\documentclass[aps,prb,amsmath,twocolumn,amssymb,titlepage,superscriptaddress,showpacs,floatfix]{revtex4-1}
\usepackage[english]{babel}
\usepackage{graphicx}
\usepackage{subcaption}
\usepackage{cleveref}
\usepackage{transparent}
\usepackage{amsmath}
\usepackage{amssymb}
\usepackage{epstopdf}
\usepackage{amsfonts}
\usepackage{bm}
\usepackage{mathdots}
\usepackage{times}
\usepackage{pdfpages}
\usepackage{mathtools}
\usepackage{stmaryrd}
\usepackage{tabularx}
\usepackage{hhline}
\usepackage{lipsum}
\usepackage{empheq}

\makeatletter
\AtBeginDocument{\let\LS@rot\@undefined}
\makeatother

\newcommand*{\citen}[1]{%
  \begingroup
    \romannumeral-`\x 
    \setcitestyle{numbers}%
    \cite{#1}%
  \endgroup   
}

\usepackage[
   justification=raggedright,
   format=plain]{caption}

\newcommand{\ave}[1]{\langle #1 \rangle}
\newcommand{\bolds}[1]{\boldsymbol #1}

\DeclareMathOperator{\sign}{sign}

\begin{document}
\title{Pairing and non-Fermi liquid behavior in partially flat-band systems:\\
Can we go beyond the nesting physics?}

\author{Sharareh Sayyad}
\email{sharareh.sayyad@neel.cnrs.fr}
 \affiliation{Institute for Solid State Physics, University of Tokyo, Kashiwanoha,
 Kashiwa, Chiba 277-8581, Japan}
 \affiliation{University Grenoble Alpes, CNRS, Grenoble INP, Institut N\'{e}el, 38000 Grenoble, France}

 \author{Edwin W. Huang}
 \affiliation{Department of Physics, Stanford University, Stanford, California 94305, USA}
 \affiliation{Stanford Institute for Materials and Energy Sciences, SLAC National
 Accelerator Laboratory and Stanford University, Menlo Park, CA 94025, USA}
 
 \author{Motoharu Kitatani}
 \affiliation{Institute of Solid State Physics, Vienna University of Technology, A-1040 Vienna, Austria}
 
 \author{Mohammad-Sadegh Vaezi}
 \affiliation{Pasargad Institute for Advanced Innovative Solutions~(PIAIS), Tehran, Iran}
 \affiliation{Department of Physics, Washington University, St. Louis, MO 63160, USA}
 
 \author{Zohar Nussinov}
 \affiliation{Department of Physics, Washington University, St. Louis, MO 63160, USA}
  
\author{Abolhassan Vaezi}
\email{vaezi@stanford.edu}
\affiliation{Department of Physics, Stanford University, Stanford, CA 94305, USA}
\affiliation{Stanford Center for Topological Quantum Physics, Stanford University, Stanford, California 94305-4045, USA}
\affiliation{Department of Physics, Sharif University of Technology, Tehran 14588-89694, Iran}

 \author{Hideo Aoki}
 \affiliation{National Institute of Advanced Industrial Science and Technology (AIST), Tsukuba 305-8568, Japan}
 \affiliation{Department of Physics, University of Tokyo, Hongo, Tokyo 113-0033, Japan}


\begin{abstract}
While many-body effects in flat-band systems are receiving 
renewed hot interests in condensed-matter physics 
for superconducting and topological properties as well as 
for magnetism, studies have primarily been restricted to 
multiband systems (with coexisting flat and dispersive bands).  
Here we focus on {\it one-band} systems where a band is 
``partially flat" comprising 
flat and dispersive portions in k-space to reveal 
whether intriguing correlation effects can arise 
already on the simplest possible one-band level.  
For that, the two-dimensional repulsive Hubbard model is 
studied for two models having different flat areas, 
in an intermediate-coupling regime with the 
FLEX+DMFT~(the dynamical mean-field theory combined with 
the fluctuation exchange approximation).  
We have a crossover from ferromagnetic to antiferromagnetic spin fluctuations as the band filling is varied,
and this triggers, 
for the model with a wider flat portion, 
a triplet-pair superconductivity 
favored over an unusually wide filling region, 
which is taken over by a sharply growing singlet pairing.  
For the model with a narrower flat portion, 
$T_C$ against filling exhibits an unusual {\it double-peaked} Tc dome, associated with different numbers 
of nodes in the gap function having 
remarkably extended pairs in real space. We identify 
these as a manifestation of the physics outside the conventional nesting physics
where only the pair scattering across the 
Fermi surface in designated (hot) spots is relevant. 
Another correlation effect arising from the flattened band 
is found in a non-Fermi-liquid behavior as detected in the momentum distribution function,
frequency dependence of the self-energy and spectral function.  
These indicate that unusual correlation physics can indeed 
occur in flat-band systems.
\end{abstract}
\maketitle

\section{Introduction}

\begin{figure*}[t]
\includegraphics[height=0.3\textheight, width=0.88\textwidth]{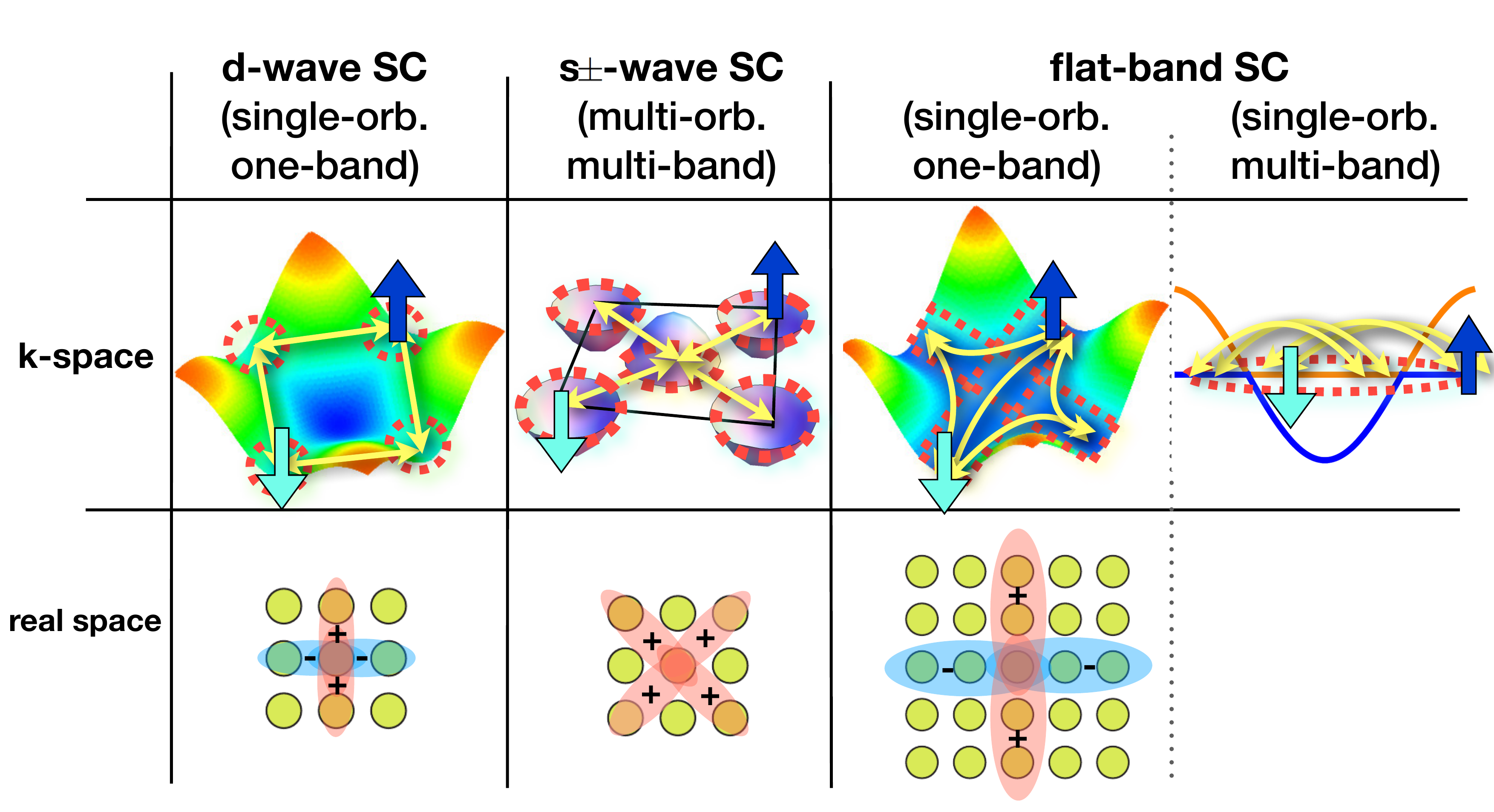}
\caption{We schematically 
compare ordinary single-orbital, one-band case (here for a d-wave 
SC; leftmost column) 
and multi-orbital, multi-band case (here for s$_{\pm}$; 
second column from left), both with 
specific ``hot spots" (dashed circles in red) across which the nesting 
vectors (yellow arrows) designate how pairs (blue and cyan arrows) hop.  
These are contrasted with flat-band systems for 
single-orbital, one-band case (second from right) and 
single-orbital, multi-band case (rightmost). The top row 
depicts k-space, while the bottom row displays pairs 
in real space~\cite{Concepfig}. The pairing for the multi-band case~\cite{Kuroki2005}
is an inter-band $s_{\pm}$, which is 
difficult to represent in real space.
}
\label{Fig:concband}
\end{figure*}

While there is a long history for the study of flat-band 
systems as initiated by interests in ferromagnetism~\cite{Lieb1989, Mielke1991, Tasaki1992, Tanaka2003, Katsura2010}, 
there is a recent surge of interests in flat-band superconductivity, 
where possibilities are explored for unconventional superconductivity 
favored by the 
flat-band structure~\cite{Kuroki2005, Tovmasyan2013, Kobayashi2016, Kauppila2016}.  
As for attractive electron-electron interactions, T\"{o}rm\"{a}'s group has shown that a flat band
can indeed favor superconductivity when the band is topological, with the superfluid weight lower-bounded
by the topological number~\cite{Tovmasyan2016, Risto2018, Tovmasyan2018, Liang2017, Liang2017b}.  
For repulsive interactions,
on the other hand, a key question is how the presence of flat bands affects electron correlation processes.  
In repulsively-interacting flat-band systems, 
spin alignment tends to lower the total energy due to unorthogonalizable Wannier orbitals through
Pauli's exclusion principle~\cite{Tasaki1992, Kobayashi2016}. 
For unconventional superconductivity, gap functions for both copper- and iron-based superconductors,
respectively with d and s$\pm$ pairings, are maximized by the 
pair scattering processes with specific momentum 
transfers~(see Fig.~\ref{Fig:concband}, top left), where 
the spin fluctuations with these wave vectors glue electrons with opposite spins~\cite{Hosono2015}.
For systems having a flat band coexisting with dispersive band(s), it has been suggested
that a key process is the quantum-mechanical virtual 
hopping of Cooper pairs between the flat and dispersive 
bands mediated by spin fluctuations arising from the 
repulsive interaction~\cite{Kuroki2005, Kobayashi2016, Matsumoto2018} 
(see Fig.~\ref{Fig:concband}, top right).  
There, it is noticed that an optimum situation is when the Fermi energy is close to, but away from, the flat band,
where the virtual pair scattering still occurs. In other words, the flat band in this situation is 
``incipient"~\cite{Qian2011}. 
There, one intriguing observation is that 
the pairing, as detected from the density matrix 
renormalization group (DMRG), involves large entanglement 
when the flat band is topological~\cite{Kobayashi2016}.

These proposals for the flat-band magnetism and 
superconductivity have so far focused on multi-band systems, as 
exemplified by Lieb's, Mielke's and 
Tasaki's models, where one of the multi-bands is 
flat while other(s) are dispersive.  
Now, a fundamental question is: can 
interesting strong-correlation phenomena such as high Tc 
superconductivity occur in simpler 
{\it one-band} systems that 
have flat {\it portion(s)} in the dispersion in the 
momentum space?  This is an interesting possibility, since, 
even when the Fermi energy resides 
on the dispersive part, quantum states are expected to be significantly altered
through the virtual pair-scattering processes between the flat and dispersive
portions of the band as well as the pair-scatterings within the flat region, both with many channels 
(which turns out to be allowed due to partial occupation of the flat portion
caused by correlation effects as we shall show; see Fig.~\ref{Fig:band}).
This will be outside the conventional ``nesting physics" for  dispersive bands where the processes
occur on Fermi surfaces.  
Thus it is intriguing whether the one-band case can be as good as, or even better than, the multi-band case.  
Motivated by these intuitions, here we explore two different flat-band models, 
where we start with a tight-binding 
(``$t$-$t'$") model with nearest and second-neighbor hoppings.  
By controlling them, we have large flat regions in the 
dispersion with the vanishingly small group velocity.
In the second model, we truncate the dispersion below certain energy into a
flat one to single out the effect of the flat part.  
Since the density of one-electron states diverges in these flat regions, 
perturbative approaches, e.g., the Schrieffer-Wolff transformation~\cite{Schrieffer1966},
might fail even in the weak electron-electron interaction regime.  
In Ref.\citen{Huang2019} the truncated model is examined 
where the unbiased determinantal quantum Monte Carlo method (DQMC)~\cite{White1989,White1989b} 
is used to show a 
Mott-insulating physics for a repulsive interaction and 
enhanced superconductivity for an attractive interaction 
in the weak-coupling regime and at intermediate temperatures, 
whereas the present paper addresses superconductivity for {\it repulsive} 
interactions.  
The flat portion also poses an interesting question of whether non-Fermi
liquid behavior can arise due to the flatness.  

Thus the purpose of the present work is to look into superconducting 
and non-Fermi liquid properties upon varying the band filling.
For that, we adopt here, along with the DQMC method,
the FLEX+DMFT method~\cite{Kitatani2015, Gukelberger2015, Kitatani2017}
which is a combination of the dynamical mean-field theory~(DMFT)~\cite{Metzner1989, Georges1992, Georges1996}
and the fluctuation-exchange approximation~(FLEX)~\cite{Bickers1989, Scalapino1989}. 
The DQMC is a numerically exact method but is applicable for limited parameters.
The FLEX+DMFT is a diagrammatic approximation and can deal with Mott's insulation for strong coupling,
but here we focus on an intermediate coupling regime. We shall show that magnetism exhibits 
a dominant ferromagnetic spin correlation at small band fillings, 
which crosses over to antiferromagnetic spin structures toward half-filling.
This concomitantly dominates superconductivity, where the pairing symmetry 
is found to change from spin-triplet to singlet.  Remarkably, 
the gap function sensitively depends on the Fermi energy sitting around the boundary between 
the flat and dispersive parts in such a way that 
(i) for the truncated model with a wider flat portion, this triggers 
a triplet pairing favored over an unusually wide filling region, 
which is taken over by a sharply growing singlet pairing toward half-filling.
(ii) For the $t$-$t'$ model with a narrower flat portion, 
$T_C$ against filling exhibits an unusual {\it double-peaked} 
$T_C$ dome associated with different numbers of nodes in the gap function.  
The unusually large numbers 
of nodal lines exhibit significantly extended pairs in real space in both models.  
Since these come from pair scatterings that involve the flat portions, 
we shall identify them as a manifestation of the physics outside the conventional nesting physics 
(with only the pair scattering across Fermi surface in designated (hot) spots being relevant).  
We shall further reveal that a non-Fermi liquid behavior 
arises as detected in various observables such as a momentum distribution function that
is fractional over the flat region, 
and the self-energy with a fractional-power-law frequency dependence accompanied by a characteristic spectral 
function.  
Thus we shall conclude that partially-flat band systems can indeed harbor quite different
and versatile physics from the ordinary bands.

\section{Model and Methods}

We consider the repulsive Hubbard model on the square lattice,
\begin{equation}
 H=
 \sum\limits_{\bolds{k}\sigma} \varepsilon_{\bolds{k}} c^{\dagger}_{\bolds{k}\sigma} c_{\bolds{k}\sigma}
  + U \sum\limits_{i} n_{i \uparrow} n_{i \downarrow}
  - \mu \sum\limits_{i\sigma} n_{i \sigma},
\end{equation}
where $c^{\dagger}_{\bolds{k}\sigma}$ creates an electron with spin $\sigma$ and
momentum $\bolds{k}$, $\varepsilon_{\bolds{k}}$ is the noninteracting band dispersion, 
$n_{i \sigma}=c^{\dagger}_{i\sigma}c_{i \sigma}$, 
$U (>0)$ is the repulsive on-site interaction, and $\mu$ is the chemical potential.

\begin{figure}[t]
\includegraphics[width=0.46\textwidth]{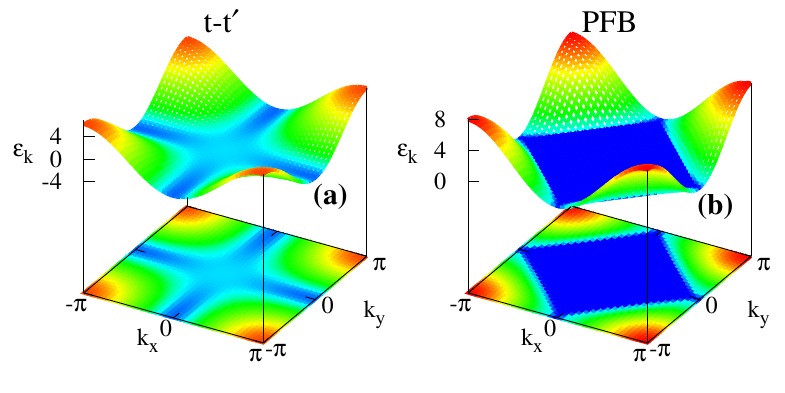}
\caption{One-electron band dispersions for the $t$-$t'$~(a) and PFB~(b) models. 
Blue region in (b) represents $\varepsilon_{\bolds{k}} =0$. 
}
\label{Fig:band}
\end{figure}

Here we consider two models (Fig.~\ref{Fig:band}): 
the first one is the $t$-$t'$ model on a square lattice 
with the 
nearest-neighbor ($t$) and the second-neighbor ($t'$) hoppings with a 
dispersion, 
\begin{align}
 \varepsilon_{\bolds{k}}^{t{\rm-}t'} = -2 t [\cos(k_{x}) + \cos(k_{y})] 
   -4t' \cos(k_{x})\cos(k_{y}).
\end{align}
If we set $t'\simeq -t/2$ we can flatten the dispersion along $\Gamma$-M lines 
(with $t'=-0.548t$ here for minimizing the curvature) as displayed in Fig.~\ref{Fig:band}(a).

The second model has a dispersion truncated as
\begin{align}
 \varepsilon_{\bolds{k}}^{\rm PFB}=
 \left[
 1+ {\cal F} \sign(\varepsilon_{\bolds{k}}^{\rm cosine})\right]
 \varepsilon_{\bolds{k}}^{\rm cosine},
\end{align}
to have a perfectly flat bottom (ideal ``partially-flat band"; PFB).  
Here $\varepsilon_{\bolds{k}}^{\rm cosine} \equiv -2 t [\cos(k_{x}) + \cos(k_{y})]$ 
is the cosine-band for the nearest-neighbor hopping model, 
and the parameter ${\cal F}$ controls the truncation, e.g., for ${\cal F}=1$ the negative-energy 
part of the cosine band is flattened as displayed in Fig.~\ref{Fig:band}(b).  
To have the same total band width ($=8$) as the cosine band, 
we set $\varepsilon_{\bolds{k}}^{\rm PFB}=2\varepsilon_{\bolds{k}}^{\rm cosine}$
for the positive part when ${\cal F}=1$. 
In this paper, we take $t$ as the unit of energy.

As for the band filling, $\ave{n}=\ave{n_{\uparrow}} +\ave{n_{\downarrow}}$, 
the noninteracting Fermi energy lies close to the flat region for $\ave{n}\lesssim1$. 
Here we study paramagnetic phases with no spin imbalance, basically with FLEX+DMFT. 
In FLEX+DMFT, the local self-energy is obtained from the DMFT procedure,
with the FLEX local self-energy subtracted to avoid double counting in a 
double self-consistent loop\cite{Kitatani2015}. 
As an impurity solver for the DMFT,
we adopt the modified iterative perturbation theory~\cite{Arsenault2012, Kajueter1996}. 
With FLEX+DMFT we do not address here the strong-coupling regime for $U$ exceeding
the bandwidth, nor very dilute fillings for convergence reasons. 
For $U$ greater than the bandwidth, employing  the continuous-time quantum Monte Carlo~\cite{Gull2011},
or the one-crossing approximation~\cite{Vildosola2015} as the impurity
solver incorporates dynamical vertex corrections more properly but,
independent of the impurity solver, our FLEX+DMFT formalism suffers from a lack of spatial vertex corrections.
To sanity-check and benchmark our FLEX+DMFT results, we compare them
with DQMC results at relatively high temperatures where the sign problem is less severe.
More precisely, in the DQMC, 
the fermionic sign problem makes the accessible temperature~($T$) for $U\leq 2$
restricted to $T\leq U/15$, see also Ref.~\citen{Huang2019}, while we can go to lower $T$s in FLEX+DMFT.
DQMC simulations are performed on a $16 \times 16$ periodic cluster,
while FLEX+DMFT is performed for a $64 \times 64$ momentum grid.

\section{Results}

\begin{figure}[t]
\includegraphics[width=0.45\textwidth]{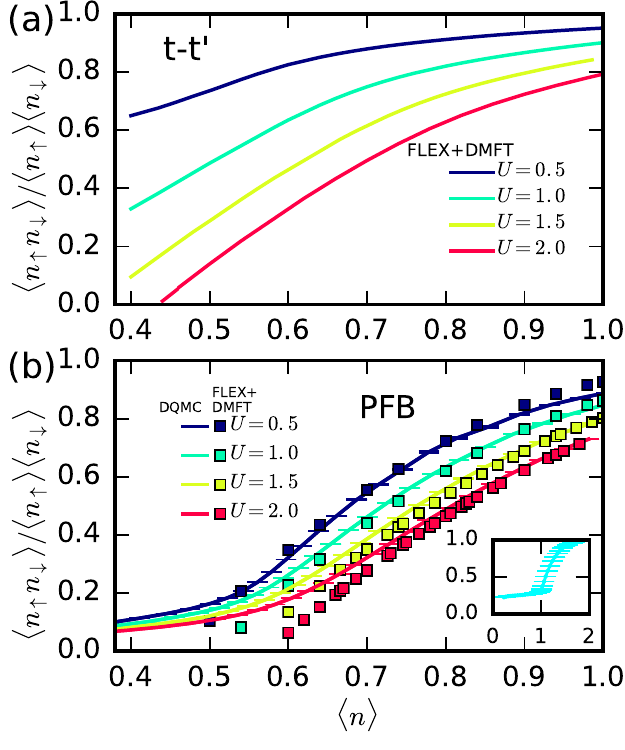}
\caption{
Double occupancy (normalized by the uncorrelated 
value~($\ave{n_{\uparrow} }\ave{n_{\downarrow}} =\ave{n}^{2}/4$)) against the band filling $\ave{n}$
for the $t$-$t'$~[solid curves in (a)] and iPFB~[squares in (b)] models, obtained with the FLEX+DMFT.  
Solid curves with error bars in (b) represent DQMC result for the PFB model~\cite{Huang2019}.  
The results are for $U= 0.5 - 2.0$ 
at temperature $T=U/15$. 
Inset in (b) is the double occupancy for the cosine band for $U=12$ at the same temperature with DQMC.
Error bars are determined by jackknife resampling.
}
\label{Fig:docc}
\end{figure}

Let us start with the double occupancy of electrons against the band filling in Fig.~\ref{Fig:docc}.  We first 
recall that, for the ordinary cosine band, the double occupancy starts to 
increase in the strong-coupling regime and above half-filling ($\ave{n}>1$) 
where the number of electrons exceeds the number of lattice sites, see inset of Fig.3 and also Ref.~\citen{Huang2019}.
If we first look at the result for the PFB model, we can see 
quite a different behavior, where the double occupancy 
starts to grow already around $\simeq 0.6$ well below half-filling, in 
both FLEX+DMFT [squares in Fig.~\ref{Fig:docc}~(b)] and DQMC results (solid curves). 
We can particularly note that even at a very weak $U=0.5$ the double occupancy arises
when significantly less than half-filled ($\ave{n}\gtrsim 0.6$), which can only occur 
in the cosine band above $\ave{n}\simeq 1$ 
at strong $U \gg \text{bandwidth}$.  
Thus we deduce that the flat region 
makes the weak interaction sufficient for the emergence of
 the  correlation effect. To endorse this, 
we turn to the double occupancy for the $t$-$t'$ model 
obtained with FLEX+DMFT [solid curves in Fig.~\ref{Fig:docc}~(a)]. 
We again encounter the double occupancy well below the half-filling.  
The double occupancy in the $t$-$t'$ model is 
greater than in the PFB, which is understandable since the flat region in the former is much narrower.

\begin{figure}[t]
\includegraphics[width=0.45\textwidth]{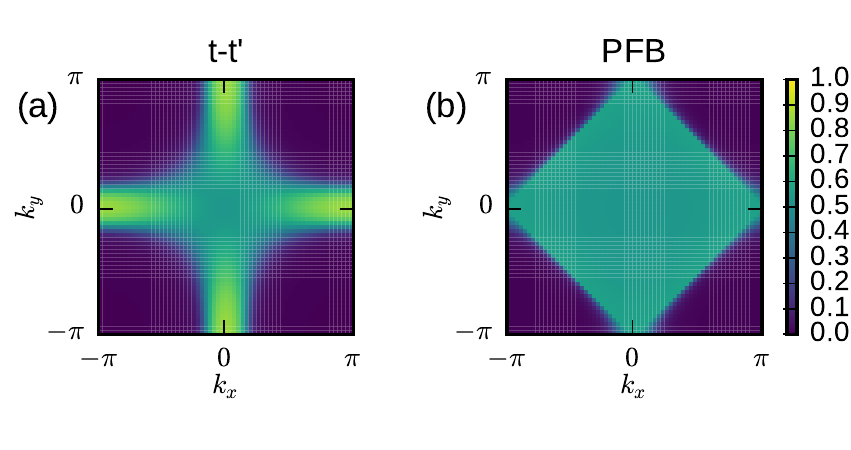}
\caption{
(a) Momentum-dependent distribution function $n_{\bolds{k}}$ for the $t$-$t'$ model at $\ave{n}=0.5$.  
(b) The same 
for the PFB model at $\ave{n}=0.62$. 
Results are computed with FLEX+DMFT on a $64\times 64$ momentum grid.  We have $U=2$ and
an inverse temperature~$\beta=7.5$ for both results.
}
\label{Fig:nk}
\end{figure}

We can then examine 
the electron configuration in the momentum space 
to compare between the flat 
and dispersive parts (i.e., how electrons doubly-occupy the flat regions before the dispersive regions are filled).  
Figure~\ref{Fig:nk} presents the momentum-dependent distribution
function~$n_{\bolds{k}}=\frac{1}{2} \sum_{\sigma} \ave{c^{\dagger}_{\bolds{k}\sigma} c_{\bolds{k}\sigma}}$,
where panel (a) is for the $t$-$t'$ model at a filling $\ave{n}=0.5$, while (b)
is for the PFB at $\ave{n}=0.62$, both for $U=2$.  
The chosen fillings are respectively around the fillings at which the double occupancy
starts to rise in Fig.~\ref{Fig:docc}.  
The figure is obtained with FLEX+DMFT, but we again observe a qualitative agreement between DQMC and FLEX+DMFT
results~(see Appendix~\ref{App:comp}).    
For the $t$-$t'$ model at $\ave{n}\simeq 0.5$, the occupation in the 
flat region along $k_x=0$ and $k_y=0$ is 
close to, but smaller than, 
unity with $0.7<n_{\bolds{k}}<0.85$.  
For the PFB model at $\ave{n}=0.62$, 
we can see an almost constant and half-filled $0.52<n_{\bolds{k}}<0.55$ over the flat region
bounded by $|k_{x}|+|k_{y}|\leq \pi$ in that model.  
Larger occupation in the $t$-$t'$ model should again be related to its narrower flat region.  

\begin{figure}[t]
\includegraphics{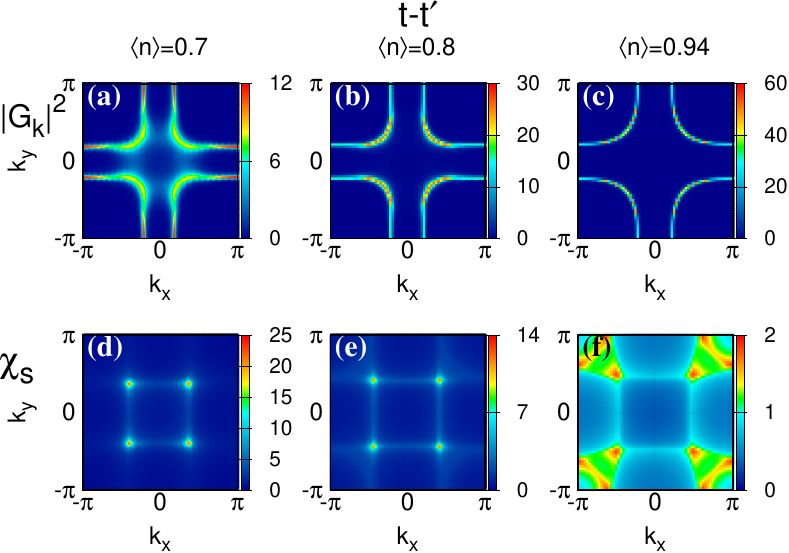}
\caption{For the $t$-$t'$ model, 
Green's functions (top panels) and spin susceptibilities (bottom) 
are color-coded in 
momentum space for fillings $\ave{n}=0.7$~(a, d), $0.8$~(b, e), and $0.94$~(c, f).
All the results are for $U=3$, $\beta=33$, 
but note different color codes for different panels.
}
\label{Fig:chisTHB}
\end{figure}

\begin{figure}[t]
\includegraphics{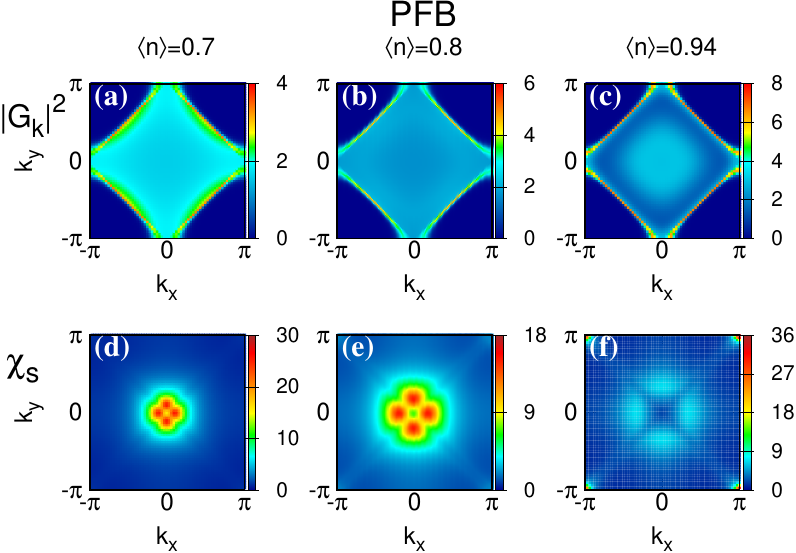}
\caption{
The same as in the previous figure for the PFB model.
In panel (f) maxima exist at $(\pm \pi, \pm \pi)$. 
}
\label{Fig:chisiPFB}
\end{figure}

The above results show that the electrons are selectively {\it crammed} 
into the flat portion causing double occupation before the dispersive portion starts to be occupied. 
This would not be surprising since the flat portions are 
situated at lower energies, but a remarkable point is the following: 
(i) The occupation is fractional, somewhere between 
the single and double occupations, and 
(ii) the occupation occurs all over the 
flat portions with basically the same occupied area as we vary the total 
band filling (compare Fig.\ref{Fig:nk} with 
Fig.\ref{Fig:nk_nc} in Appendix~\ref{App:3/4occupation}) 
in both models.  In this sense Luttinger's theorem~\cite{Kridsanaphong2018} 
does not seem to apply here.   
To explore the Fermi surface formation, we plot the Green's function for both models
in Figs.~\ref{Fig:chisTHB},\ref{Fig:chisiPFB} (top panels), where 
sharp peaks would define the Fermi surfaces. 
While the Fermi surfaces are visible in the $t$-$t'$ model 
(Fig.~\ref{Fig:chisTHB}), they are not very well-defined 
for the PFB model (Fig.\ref{Fig:chisiPFB}). 
We come back to this point below in terms of the frequency dependence of the self-energy. 

Let us now turn to the spin structure against the band filling.   
The static spin correlation function, 
$\chi_{s}(\bolds{k})=2\int_{0}^{\beta} {\rm d}\tau \ave{S^{z}_{\bolds{k}}(\tau)S^{z}_{-\bolds{k}}(0)}$~\cite{Roshen1983}
is displayed for the $t$-$t'$ ~(Fig.~\ref{Fig:chisTHB}) and PFB~(Fig.~\ref{Fig:chisiPFB}) models 
at $U=3$, $\beta\equiv 1/(k_BT)=33$ and $\ave{n}=0.7 - 0.94$. 
Overall, the spin correlation is seen to be large over 
{\it streaks} or wide {\it plateaus} (rather than usual spots),
which should come from the flattened bands.  
More precisely, 
reflecting the structure of Green's function, 
the $t$-$t'$ model shows streaks across some mid-points in the Brillouin zone, which cross over to wider and 
more complex structures as we approach $\ave{n}=1$.  
A smaller overall value of the spin susceptibility in Fig.~\ref{Fig:chisTHB}(f) may be attributed to 
coexistence of spin fluctuations coming from 
occupied flat and dispersive portions.  
The PFB model, on the other hand, shows a crossover from ferromagnetic 
spin fluctuations, which is expected as 
in the spin alignment in the half-filled 
flat branch in multi-band models, to wider 
plateaus with peaks shifting away from $\Gamma$ point, and 
finally to antiferromagnetic spin fluctuations 
with peaks around $(\pm \pi,\pm \pi)$ as we approach $\ave{n}=1$.
As for the charge susceptibility,
~$\chi_{c}(\bolds{q})=\int_{0}^{\beta} {\rm d}\tau \ave{n_{\bolds{q}}(\tau)n_{-\bolds{q}}(0)}$,
we observe a similar trend in both models, but $\chi_{c}$ 
is an order of magnitude smaller than the spin susceptibility. 
We shall see below that the spin structure governs the structures of the self-energy, local spectral function as well 
as pairing.

\begin{figure}[b]
\includegraphics[width=0.48\textwidth]{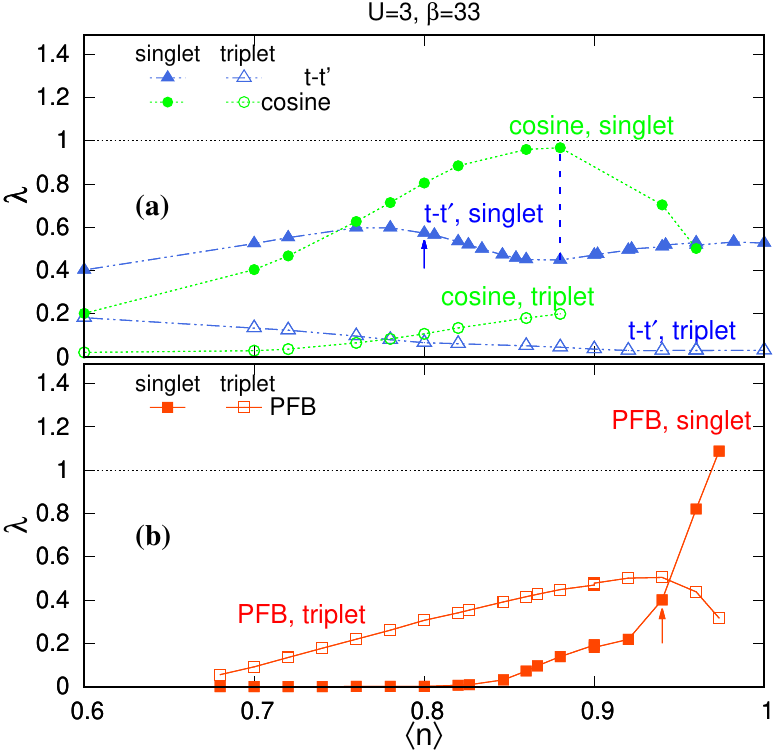}
\caption{
Largest eigenvalue $\lambda$ of the Eliashberg equation versus filling
for the singlet~(filled symbols) and triplet~(empty) pairings for $t$-$t'$~(triangles)~(a), cosine-band~(circles)~(a),
and PFB~(squares)~(b) models for $U=3$, $\beta=33$. Arrows indicate the $\langle n \rangle_{\rm c}$ at 
which the spin susceptibility is peaked for 
respective models, while the dashed vertical line indicates a change in the pairing symmetry, see text.
}
\label{Fig:lambda}
\end{figure}

Now we are in position to explore the superconducting phases 
with the linearized Eliashberg equation for 
the gap function $\Delta$,
\begin{equation}
 \lambda \Delta(k)
 =
 -\frac{1}{\beta} \sum\limits_{k'}
 V_{\rm eff}(k-k')
 G(k')
 G(-k')
 \Delta(k'),\label{eq:eliashberg}
\end{equation}
where $\lambda$ is the eigenvalue, 
$k \equiv (\bolds{k}, i\omega_{n})$ with $\omega_{n}$ being the Matsubara frequency with $\sum_{\bolds{k}}=1$, 
$V_{\rm eff}= U + 3 U^{2} \chi_{s}/2-U^{2} \chi_{c}/2$ 
is the effective pairing interaction, and $G$ is Green's function.  
The eigenvalue is a measure of superconducting instabilities with $\lambda=1$ 
marking $T_C$.  
Figure~\ref{Fig:lambda}, a key result of this work, plots $\lambda$ 
for singlet~(filled symbols) and triplet~(empty) pairings for the $t$-$t'$ and PFB models. 

\begin{figure}[t]
\includegraphics{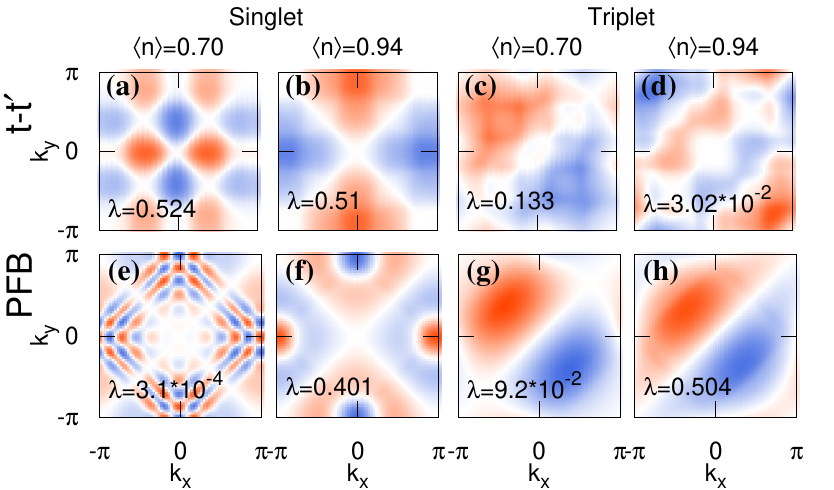}
\caption{
Singlet (left four panels) and triplet (right four) 
gap functions in momentum space for the $t$-$t'$~(first row) and PFB~(second row) models.  
Filling is $\ave{n}=0.7$~(a,c,e,g) or $0.94$~(b,d,f,h). 
For each of the triplet cases, 
another one rotated by 90 degrees is degenerate due to 
the tetragonal symmetry of the lattice.
Maximum eigenvalue of the Eliashberg equation is indicated in each panel.   
Color code for the gap function 
is bluish~(reddish) for negative~(positive) values, 
for which we have omitted the color bars since the 
linearized Eliashberg equation does not indicate magnitudes of $\Delta$.  
All the results are for $U=3$, $\beta=33$.
Note that the $\lambda$ in (d,e,g) 
is vanishingly small ($< 10^{-1}$), so should not be taken seriously.
}
\label{Fig:gap_k}
\end{figure}

If we first look at the result for the PFB model, triplet pairing is 
favored with larger $\lambda$s over a remarkably wide 
region of the filling, which indicates the importance of the wide flat region accompanied by ferromagnetic 
fluctuations.  Then a singlet pairing 
rapidly dominates as we approach $\ave{n}=1$.  
In the $t$-$t'$ model, with a narrower flat portion and 
associated spin fluctuations (Fig.~\ref{Fig:chisTHB}), 
singlet pairing dominates over the whole region studied here, 
but with a curious {\it double-dome} structure in $T_C$.  
Both of these are in dramatic contrast 
with the usual cosine-like bands, where the singlet d-wave 
pairing dominates with a single dome in $\lambda$ 
around $\ave{n} \approx 0.9$~\cite{Kitatani2015, Kitatani2017, Kitatani2019}.
The sharp enhancement in the singlet pairing 
close to the half-filling in the PFB, which should come 
from the prevailing antiferromagnetic fluctuations, 
has $\lambda$ that is larger than $t$-$t'$ and 
even the cosine-band counterparts.  
We note that  this takeover 
(an arrow in Fig.~\ref{Fig:lambda}~(b)) occurs when the flat-band filling exceeds about $3/4$
(see Appendix~\ref{App:3/4occupation}), which in fact coincides with a critical filling, $\ave{n}_{\rm c}$, 
at which the DMFT spin susceptibility is peaked (see Appendix~\ref{App:chis}),
and the exponent in the self-energy is also peaked as we shall see in Fig.~\ref{Fig:self-energy}(b) below.  
As for the singlet pairing in the $t$-$t'$ model, 
we can see that the left peak in $\lambda$ occurs around $\ave{n}_{\rm c}$ for this model.

\begin{figure}[t]
\includegraphics{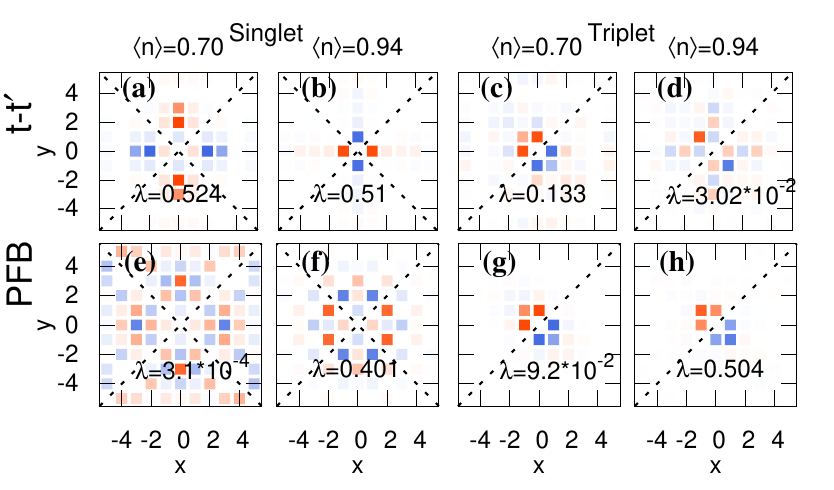}
\caption{
The same as in the previous figure in real space.  
Dashed lines represent nodes. 
}
\label{Fig:gap_r}
\end{figure}

So let us now fathom these results in terms of 
filling-dependent singlet and triplet gap functions 
in momentum space in Fig.~\ref{Fig:gap_k}, 
or in real space in Fig.\ref{Fig:gap_r}, for the $t$-$t'$ and PFB models. 
We can immediately see that all the gap functions are anisotropic and possess
nodal lines whose number sensitively depends on the filling. 
In the $t$-$t'$ model, cases similar to the 
usual d-wave [$\Delta(\bolds{k}) \sim 
\cos(k_{x}) - \cos(k_{y})$] exist [as in Fig.~\ref{Fig:gap_k}(b), 
Fig.\ref{Fig:gap_r}(b)], but more generally admits structures, 
$\Delta_{\rm singlet}(\bolds{k}) \sim \cos(\gamma k_{x}) - \cos(\gamma k_{y})$, 
where $\gamma (=1, 2, \cdots)$ characterizes the number of nodal lines.  
For instance, we have $\gamma= 1\rightarrow2$ 
for $\ave{n}=0.94\rightarrow0.7$ in the $t$-$t'$ model. 
This shows that, as we go away from the half-filing at which antiferromagnetic fluctuations dominate, the usual 
$d_{x^2-y^2}$ wave changes into something more complicated. 

If we turn to the gap functions in real space in Fig.\ref{Fig:gap_r}, we can realize 
that the larger the number of nodal lines, the more extended the pairs over several lattice spacings in real space. 
Similar long-range pairings have also been explored for 
quasi-one-dimensional and 2D systems \cite{Tanuma2002,Kuroki2004}, where each pair becomes
more spatially extended as we go from p-wave to d and f with the number of nodes increasing.  
For the triplet gap function~\cite{arita2004} in Fig.\ref{Fig:gap_k}~(c,d,g,h),
we also tend to have unusually extended pairing with larger numbers of nodes. In the literature, the random-phase 
approximation~(RPA) has been used to obtain 
the filling-dependent gap symmetry in the $t$-$t'$ model~\cite{Romer2015}, but the present results exhibit 
different bahavior such as an absence of s-waves seen 
in RPA, which should be due to the self-energy effects incorporated more accurately here.  
For the PFB model, triplet gap functions are close to a simple p-wave, 
$\Delta_{\rm triplet}(\bolds{k}) \sim \sin(k_{x}) \pm \sin(k_{y})$, but extra nodes are visible.

If we go back to Fig.\ref{Fig:lambda}, the Eliashberg $\lambda$ in our partially flat-band systems 
can be smaller than those for the ordinary cosine band, 
which may be related to less compact pairing in the 
former, but 
we do have important effects peculiar to the flat-band cases: 
For the PFB, 
(i) the singlet $\lambda$ sharply blows up toward 
the filling $n=1$, and (ii) before this occurs the triplet pairing are favored over an unusually wide region of $n$.
For $t$-$t'$, (iii) 
the peculiar double-peak structure arises from 
a change in the number of nodes around the dip of $\lambda$ (marked with a 
vertical dashed line in Fig.\ref{Fig:lambda}~(a)).  
If we compare the single-band and multi-band 
flat-band systems~\cite{Misumi2017}, the former 
sharply contrasts with the behavior of the multi-band 
case\cite{Matsumoto2018} in which the Eliashberg $\lambda$ 
is shown to have a sharp dip when $E_F$ becomes too close to the 
flat band, but this does not occur in the 
present single-band case.


\begin{figure}[t]
\includegraphics{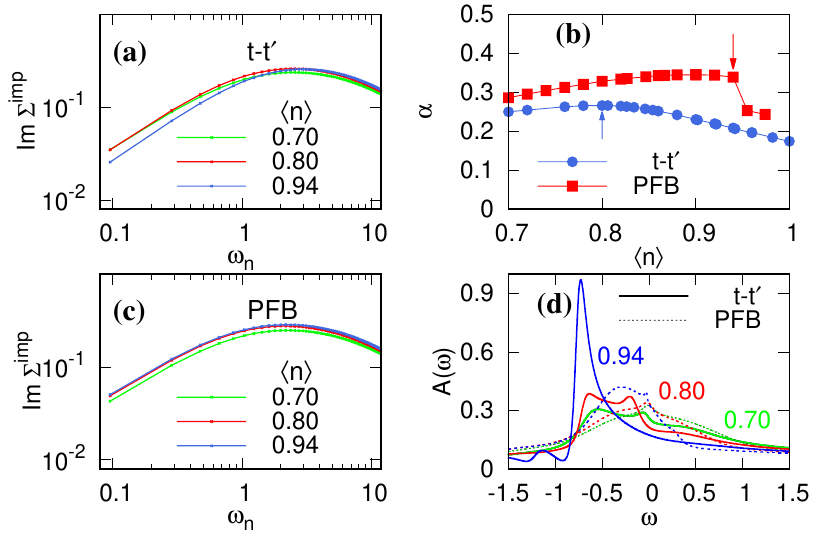}
\caption{
Imaginary part of the DMFT impurity self-energy, 
$\Sigma_{\rm imp}$, 
against Matsubara frequency, $\omega_{n}$, for $\ave{n} = 
0.7$ (green), $0.9$~(red), $0.94$~(blue) in the 
$t$-$t'$ (a) and PFB (c) models.  
(b) The exponent $\alpha$ in the fit $|{\rm Im} 
\Sigma_{\rm imp}(i \omega_{n})| \propto \omega_{n}^{\alpha}$, 
for the $t$-$t'$~(blue circles) and PFB~(red 
squares) models.  Arrows mark respective 
$\ave{n}_{\rm c}$ for the spin susceptibility peaks, which 
are seen here to coincide with the peaks in $\alpha$. 
(d) Local spectral functions in the $t$-$t'$~(solid lines) and PFB~(dotted) 
models for the band filling $\ave{n} = 0.7, 0.8, 0.94$.
All the results are for $U=3$ and $\beta=33$.
}
\label{Fig:self-energy}
\end{figure}

{\it Fermi-liquid properties}  
We finally look into the Fermi-liquid properties.  
In Fig.~\ref{Fig:self-energy}(a,c) we plot the imaginary part of the DMFT
self-energies against Matsubara frequency in the $t$-$t'$~(a)
and PFB~(c) models, for band filling 
$\ave{n}= 0.7-1.0$  at $U=3$, $\beta=33$. 
We notice that the self-energy exhibits 
a peculiar frequency dependence.  We can actually 
quantify non-Fermi liquid behavior by 
fitting the imaginary part of the self-energy to 
\[
|{\rm Im} \Sigma_{\rm imp}(i \omega_{n})|\propto \omega_{n}^{\alpha},
\]
for low Matsubara frequencies. Then 
$\alpha=1$ characterizes the Fermi liquids, while 
$\alpha<0.5$ will signify a ``bad-metallic" behavior~\cite{Werner2008,Ishida2010,Werner2016}.  Here we take 
the DMFT impurity self-energy, 
$\Sigma_{\rm imp}$, since 
we want to look at the local self-energy, which we shall later compare with 
the DMFT impurity spin-susceptibility. 
We can see in Fig.~\ref{Fig:self-energy}(b) that the 
exponent $\alpha$ increases 
up to a critical filling~$\ave{n}_{\rm c}$ that 
depends whether we have the $t$-$t'$ or PFB models. 
It is notable that the $\ave{n}_{\rm c}$ for the self-energy coincide with the critical $\ave{n}_{\rm c}$ 
($=0.82$ for $t$-$t'$, $\ave{n}_{\rm c}=0.94$ for PFB at $U=3$, $\beta=33$) 
at which the spin-susceptibility for the DMFT impurity, $\chi^{\rm imp}_{s}$, has a peak in each model,
as shown in Appendix~\ref{App:chis}.  
For further increase of the filling, $\alpha$ starts to decrease.  
The value $\ave{n}_{\rm c}=0.94$ in the PFB 
corresponds to the filling at which the flat part of the band is about $3/4$-filled, namely
we have $0.71<n_{\bolds{k}}<0.83$  on the flat portion in 
the momentum-dependent distribution function~(see Appendix~\ref{App:3/4occupation}).  
A bad-metallic behavior also appears as deformations in the local spectral functions in
Fig.~\ref{Fig:self-energy}(d), obtained via analytic continuation with the Pad\'{e} approximation. 
In particular, the local spectral functions undergo large changes for $\ave{n}>\ave{n}_{\rm c}$
with an emergence of multi-peaks that are separated by $\omega \ll U$,
see also Appendix~\ref{App:Awk}.

\section{Conclusion and discussions }
To summarize, we have studied two partially flat-band 
models ($t$-$t'$ and PFB) 
to reveal that a manifestation of the flat portion in the band gives 
a dramatic difference from the 
ordinary band to produce a peculiar sequence of the dominant pairing symmetries.  
This occurs in both models, in a manner that is 
dominated by the size of the flat region.  
For PFB with a wide flat area, triplet pairings are favored 
over a wide filling region, while for $t$-$t'$ with a narrower flat 
area, a double-dome structure in $T_C$ emerges associated 
with 
different numbers of nodes in the gap function.  
Concomitantly, pairings can become unusually extended in real space with large numbers of nodes.  
We have finally shown that non-Fermi-liquid like 
behavior exists in a power-law frequency dependence of the self-energy, etc.  
We identify these as a peculiar emergence of correlation effects in partially flat-band systems
that can occur even for intermediate electron-electron interactions. 

As Fig.~\ref{Fig:concband} suggests, the different 
pairings revealed here should come 
from quite different configurations of pair-scattering 
channels in the partially-flat band models: 
In regular bands the key process is 
the pair scattering specifically between the ``hot spots" (anti-nodal 
regions in the cuprates, and Fermi pockets in the 
iron-based superconductors\cite{Hosono2015}, 
respectively giving rise to the 
d and s${\pm}$ pairings), which is contrasted with the flat bands 
that have 
the whole {\it bunch} of pair-scattering channels involving, so 
to speak, an ``extended hot regions".  

This now leads us to make an 
observation: For ordinary bands, we can show, 
from a general phase-space volume argument\cite{Arita1999, Monthoux1999}, 
that the superconductivity mediated by spin fluctuations 
should work much more efficiently in 
two-dimensional (layered 2D) systems than in 3D.  
By contrast, 
the flat bands with (i) extended hot regions 
(with wide areas in k-space for large spin fluctuations), 
(ii) wide areas for large gap function amplitudes, 
and (iii) also wide areas for large Green's functions 
(which are involved in Eq.(\ref{eq:eliashberg})) 
may {\it evade} the above theorem 
to render 3D systems as good as in 2D.  This will make 
3D partially flat band systems interesting.

An important question of course is whether flat bands can enhance $T_C$. For the attractive Hubbard model,
the sign-free DQMC 
actually indicates that $T_C$ is nearly doubled when the band is flattened into PFB \cite{Huang2019}.  
A general question then is whether $T_C$ is enhanced in the repulsive model, which is an important future problem. 
For ordinary (cosine) bands, Kitatani {\it et al.} 
have used D$\Gamma$A (dynamical vertex approximation) to 
identify the vertex correction as the reason 
why $T_C (\sim 0.01t)$ in the spin-fluctuation mediated pairing 
is two orders of magnitude smaller than the starting 
electronic energy\cite{Kitatani2019}.  
It will be interesting to see whether the vertex 
correction in the flat-band systems can act to overcome 
this. In the present flat-band models, the spin susceptibility 
can have broad structures such as plateaus.  
One possible hint is that Yanase {\it et al.}\cite{Yanase2003} show that 
the vertex correction becomes significant in a 
model that has a featureless spin structure.

As for vanishing group velocity, this also occurs point-lie at van Hove singularities, 
and its effect on correlation physics has been discussed~\cite{Liu2018}, 
where topological superconductivity such as d+id wave is suggested. 
So it is intriguing to examine whether 
the present systems, where the group velocity 
vanishes in finite {\it areas} rather than at points, can 
accommodate topological superconductivity.  
The present work has shown transitions between 
different pairing symmetries (within singlets with different numbers of 
nodes in $t$-$t'$ and singlet-triplet transition in PFB).  
In fact, it is known that the boundary between different 
pairing symmetries is a promising venue for looking for time-reversal-broken topological superconductivity
\cite{Fernandes2013, Ahn2014, Oiwa2019}.

As for possible realizations of the present model, we can 
raise an example which is the $\tau$-type organic salt 
family, D$_2A_1A_y$, based on D (=P-$S$, $S$-DMEDT-TTF or EDO-$S$, $S$-DMEDT-TTF) in combination 
with anions $A$ (=AuBr$_2$, I$_3$, or IBr$_2$), studied by Papavassiliou {\it et al.}
~\cite{Papavassiliou1995, Aizawa2014},
which are two-dimensional metals in the $\tau$ crystal form. The band structure of a single layer 
of the $\tau$ phase contains a flat-bottomed band just as in the present $t$-$t'$ model.
Indeed, a checkerboard-patterned organic molecule in the layer makes its effective model 
a tight-binding system with $t' \simeq -0.5t$~\cite{Papavassiliou1995, Aizawa2014, Arita2000, putativematerial}.
The partially flat-band models proposed here require distant hopping amplitudes.  
In real materials, an organic $\tau$-type conductor, 
for instance, has been investigated experimentally, 
and distant hoppings are theoretically shown to gives rise to a partially
flat band~\cite{Papavassiliou1995, Aizawa2014, Arita2000}. 
As for inorganic materials, ruthenate superconductors~\cite{Wang2013, Autieri2014},
and some iron-chalcogenides\cite{Johnson2015, Subedi2008} have partially flat bands.  
They are multi-band systems, where competition between various pairing
symmetries~\cite{Kuroki2001, Mackenzie2003}, fractional power-law behavior
in the optical conductivity~\cite{Yin2012, Mirri2012, Lee2003}, and (anti)ferromagnetic
spin structures~\cite{Kawanaka2016, Hisashi2012} have been discussed. 
On the other hand, there is a recent upheaval of interests in the twisted bilayer graphene,
where the band structures are shown to have flat portions
on hexagonal lattices~\cite{Cao2018, Cao2018b, Koshino2018, Morell2010, Bistritzer2011, Mele2010, Trambly2010, Fang2016, Guorui2019, Moriyama2019, Codecido2019, Po2019}. 
This further highlights the need to understand partially flat bands more generically.
As for the space group we can extend the present idea on tetragonal lattices to hexagonal cases, which is underway. 

\section{Acknowledgments}
Numerical calculations were performed on the REIMS cluster and 
the ISSP Supercomputer in the University of Tokyo,
and on the Sherlock cluster at Stanford University.
Sh.S. and H.A. acknowledge a support from the ImPACT Program of
the Council for Science, Technology and Innovation, Cabinet
Office, Government of Japan (Grant No. 2015-PM12-05-01)
from JST. Sh.S. acknowledges funding from ANR-18-CE30-0001-01.
H.A. thanks Kazuhiko Kuroki for illuminating discussions, and is also supported
by JSPS KAKENHI Grant Nos. JP26247057, 17H06138, and CREST ``Topology" 
project from JST.  
E.W.H. was supported by the U.S. Department of
Energy (DOE), Office of Basic Energy Sciences, Division
of Materials Sciences and Engineering, under Contract
No. DE-AC02-76SF00515. 
M.S.V. and Z.N. acknowledge partial support by the National
Science Foundation (NSF 1411229). M.S.V. also acknowledges
the financial support from Pasargad Institute
for Advanced Innovative Solutions (PIAIS) under Supporting
Grant scheme (Project SG1-RCM1903-01).

\bibliography{Man_Flatband}

\appendix

\section{Comparison of DQMC and FLEX+DMFT results for momentum-dependent distribution functions}\label{App:comp}

\begin{figure}[htp]
\includegraphics{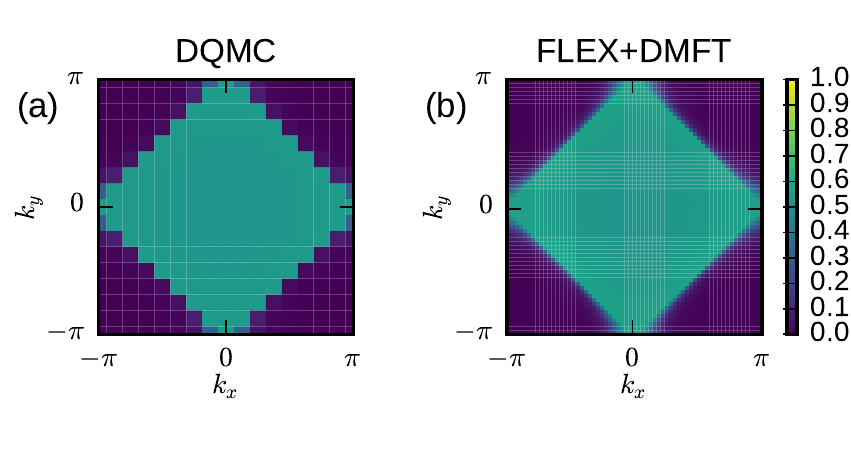}
\caption{
Momentum distribution function for the PFB model computed with DQMC on a periodic $16 \times 16$ cluster~(a), as 
compared with the result in FLEX+DMFT on a $64\times 64$ momentum grid~(b), for $U=2$,
inverse temperature~$\beta=7.5$, and filling $\ave{n}=0.62$.
}
\label{Fig:compnk}
\end{figure}

Let us here compare 
the momentum-dependent distribution functions obtained with the DQMC and FLEX+DMFT approaches for the PFB model 
at $\ave{n}=0.62$ for $U=2$, $\beta=7.5$ 
in Fig.~\ref{Fig:compnk}. The occupancy and shape of the occupied regions are seen to 
accurately agree between the two results. 
More precisely, the electron occupancy in the 
flat portions
ranges 0.52-0.55 in the DQMC~(a), and 0.52-0.57 
in the FLEX+DMFT~(b).

\section{Momentum-dependent distribution functions\\ 
at $\ave{n}_{\rm c}$ for the $t$-$t'$ and PFB models}
\label{App:3/4occupation}

\begin{figure}[htp]
 \includegraphics[width=0.47\textwidth]{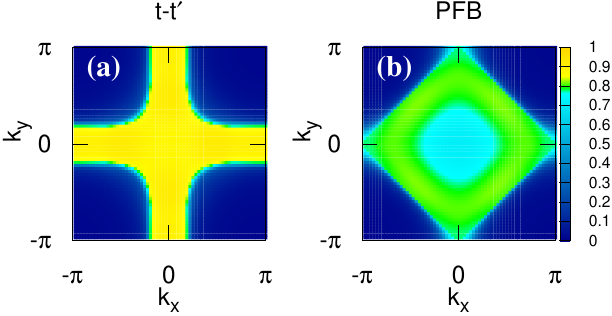}
\caption{
Momentum-dependent distribution function, 
$n_{\bolds{k}}$,  at the critical 
fillings, $\ave{n}_{\rm c}=$ 0.82 for the $t$-$t'$~(a) and 0.94 for the PFB~(b) models, for $U=3$ and $\beta=33$.
}
\label{Fig:nk_nc}
\end{figure}

We display in Fig.~\ref{Fig:nk_nc} how the momentum-dependent distribution function, $n_{\bolds{k}}$,
behaves right at the critical 
filling $\ave{n}_{\rm c}=$ 0.82 for $t$-$t'$, and 0.94 for the PFB model. 
The occupation of the flat portion of the band in the PFB system ranges from 0.71 to 0.83, i.e., about 3/4.  
In the $t$-$t'$ model, the flat portion is close to fully occupied, associated with the narrower
size of the flat region of this band.
These results should be compared with Fig.\ref{Fig:nk} in the main text, where 
the flat portion is an occupation about 1/2 in PFB.

\section{DMFT impurity spin-susceptibility}\label{App:chis}

\begin{figure}[htp]
 \includegraphics{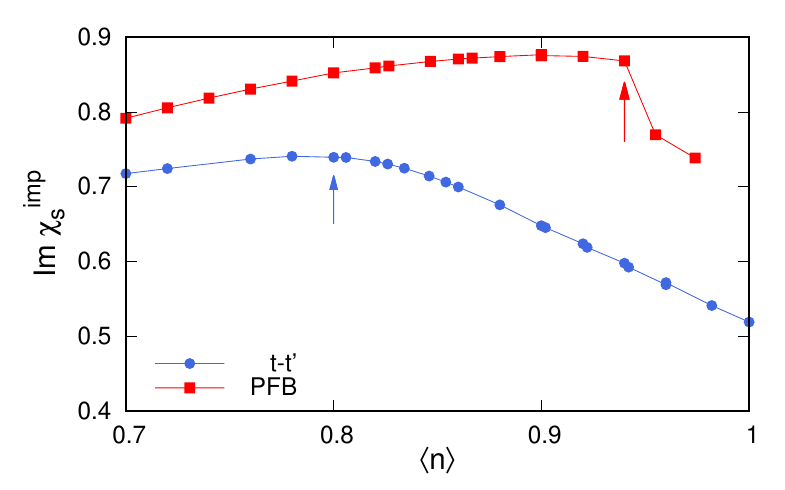}
\caption{
DMFT impurity spin-susceptibility against band filling for the $t$-$t'$~(green circles)
and PFB~(blue squares) models for $U=3$ and $\beta=33$. Arrows indicate 
the critical filling $\ave{n}_{\rm c}$ in the two models, respectively.
}
\label{Fig:loc_chis}
\end{figure}

Let us display in 
Fig.\ref{Fig:loc_chis} the DMFT spin-susceptibility~$\chi_{s}^{\rm imp}$, obtained
from the DMFT impurity Green's functions, for $U=3$ and inverse temperature $\beta=33$
in the $t$-$t'$ and PFB models. The result exhibits a 
peak (marked respectively with an arrow) in each model, which is seen to coincide 
with the critical filling $\ave{n}_{\rm c}$ for the 
self-energy behavior introduced 
in the main text, see the arrows in Fig.\ref{Fig:lambda} and Fig.\ref{Fig:self-energy}(b).

\section{Momentum-dependent spectral functions\\
at $\ave{n}_{\rm c}$}
\label{App:Awk}
\begin{figure}[htp]
\includegraphics[width=0.47\textwidth]{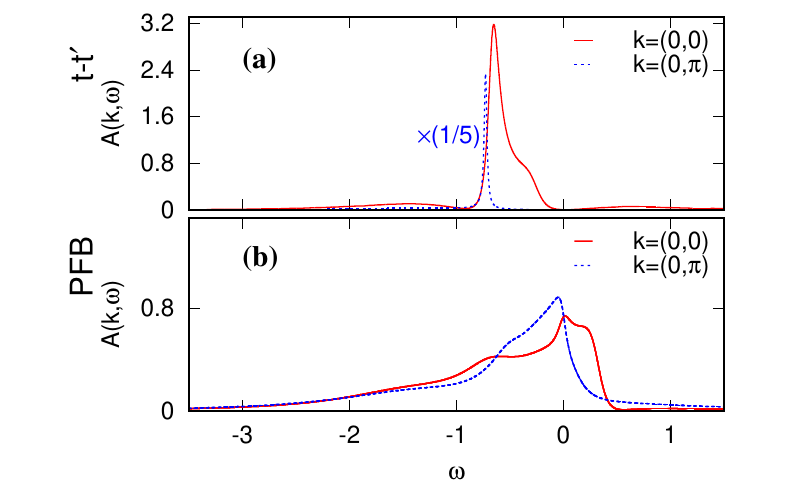}
\caption{
Spectral function, $A(\bolds{k},\omega)$, at $\Gamma$~(red solid lines) and 
X~(blue dotted) points in $t$-$t'$~(a) and PFB~(b) models, for $U=3$, $\beta=33$,
and band filling $\ave{n}=$ 0.81~(a) or $\ave{n}=$ 0.94~(b).
Spectral function at X point in (a) is devided by a factor of 5.
}
\label{Fig:Akw}
\end{figure}

We present the momentum-dependent spectral functions at $\Gamma~(0,0)$ and X~$(0,\pi)$ points in the Brillouin zone 
right at the critical band filling, 
$\ave{n}=$ 0.81 in $t$-$t'$ and 
$\ave{n}=$ 0.94 in PFB models, in Fig.~\ref{Fig:Akw}. 
The spectrum is obtained with Pad\'{e} approximation. In both panels, shoulder-like features are
seen at $\Gamma$ point as a correlation effect. 
Such a feature also appears at $(0,\pi)$ in the PFB model.
The result should be compared with the local spectral functions in Fig.~\ref{Fig:self-energy}(d) in the main text.

\end{document}